\documentclass[a4paper,superscriptaddress,twocolumn,prb,aps,showpacs,floatfix,citeautoscript]{revtex4}

\usepackage{graphicx}
\usepackage{bm}
\usepackage{amsmath,amsbsy,amsfonts,mathrsfs}
\usepackage{longtable}
\usepackage{dcolumn}
\usepackage[utf8]{inputenc}

\newcommand{\D}{{\rm d}}

\newcommand{\br}{{\bm r}}

\newcommand{\mint}[1]{\int\! \D^{3} #1 \, }
\newcommand{\mdint}[2]{\mint{#1}\!\!\!\mint{#2}}

\newcommand{\lyon}{Laboratoire de Physique de la Mati\`ere Condens\'ee et
  Nanostructures, Universit\'e Lyon I, CNRS, UMR 5586, Domaine Scientifique de
  la Doua, F-69622 Villeurbanne Cedex, France}
\newcommand{\coimbra}{Centre for Computational Physics, Department of Physics, 
  University of Coimbra, 3004-516 Coimbra, Portugal.}
\newcommand{\etsf}{European Theoretical Spectroscopy Facility.}
\newcommand{\berlin}{Institut f{\"u}r Theoretische Physik, Freie
    Universit{\"a}t Berlin, Arnimallee 14, D-14195 Berlin, Germany.}
\newcommand{\athens}{Theoretical and Physical Chemistry Institute, National Hellenic 
                     Research Foundation, Vas. Konstantinou 48, GR11635 Athens, Greece.}

\begin{document}

\title{Empirical Functionals for Reduced Density Matrix Functional Theory}

\date{\today}

\author{Miguel A.\,L. Marques}
\affiliation{\lyon}
\affiliation{\coimbra}
\affiliation{\etsf}

\author{N. N. Lathiotakis}
\affiliation{\athens}
\affiliation{\berlin}
\affiliation{\etsf}

\begin{abstract}
  We present fully empirical exchange-correlation functionals to be
  used within reduced density matrix functional theory (RDMFT). These
  are of the popular J-K form, where the function of the occupation
  numbers that multiplies the Fock orbital term is written as a Padé
  approximant. The coefficients of the Padé are optimized for a test
  set of eight molecules, and then refined for a larger set of 35
  molecules. Two different approaches were tried, either keeping the
  self-interaction terms, or by removing them explicitly from the
  functional. The functionals thus obtained involve very few
  parameters, but are able to outperform other RDMFT functionals,
  yielding correlation energies that are, on average, even slightly
  better than M{\o}ller-Plesset MP2 theory.
\end{abstract}

\maketitle

Reduced matrix density functional theory (RDMFT) has been emerging as an excellent tool for the study of correlation in
molecular systems. It is based on Gilbert's theorem\cite{gilbert}, that asserts a one-to-one correspondence between the
ground-state
many-body wave function and the one-particle reduced density matrix, $\gamma$. 
Several theoretical advantages are
immediately evident from using $\gamma$ instead of, e.g., the electronic density $\rho$ as in standard density
functional theory (DFT): i)~both the kinetic energy and the exchange energy can be trivially written as explicit
functionals of $\gamma$; ii)~consequently, the so-called correlation energy in RDMFT has a proper scaling with the
strength of the Coulomb interaction, as it is free from contamination from the kinetic term. Therefore, one could expect
that it is much easier to find good correlation functionals for RDMFT than to standard DFT.

Indeed, the past years have witnessed the appearance of several such
functionals. Most of them are of the form (for closed-shell systems)
\begin{multline}
  \label{eq:Exc}
  E_{\rm xc} = 
  -\frac{1}{2} \sum_{j,k} 
  \mdint{r}{r'} f(n_{j}, n_{k})
  \\ \times
  \frac{\varphi_{j}^*(\br)\: \varphi_{k}^*(\br')\: \varphi_{k}(\br)\: 
  \varphi_{j}(\br')}{|\br-\br'|} \,,
\end{multline}
i.e., they have the form of the usual Hartree-Fock (HF) exchange modified by the
function $ f(n_{j}, n_{k})$ of the occupation numbers. Functionals of the form
of Eq.~(\ref{eq:Exc}) are sometimes referred to as of J-K type, and involve only
Coulomb (J) and exchange (K) type integrals over the natural orbitals.

The first approximation to $E_{\rm xc}$, introduced by M\"uller in
1984,~\cite{mueller} corresponds to the formula
\begin{equation}
  f^\text{M\"uller}(n_j, n_k) = \sqrt {n_j n_k}
  \,.
\end{equation}
Later, Buijse and Baerends\cite{buijse} arrived at the same functional by modeling 
the exchange and correlation hole, while Goedecker and Umrigar (GU) considered 
a modified version by explicitly removing the self-interaction (SI) terms. 
The extremely simple form of the M\"{u}ller functional yields the correct dissociation limit of dimers of
open-shell atoms like H$_2$ (where both DFT and HF fail), but overestimates
substantially the correlation energy~\cite{staroverov,herbert} of all the
systems it has been applied to (including the
HEG~\cite{ciospernal,csanyi,nekjellium}). The GU form on the other hand, 
is more accurate in the calculation of the  correlation energy\cite{GU,staroverov,herbert},
but fails dramatically at the dissociation limit.\cite{staroverov,herbert}

Several other forms for $f(n_{j}, n_{k})$ exist in the market right now. The most precise for molecular systems seem to
be the BBC3\cite{gritsenko} functional of Gritsenko, Pernal and Baerends, and the functionals of Piris\cite{piris}. 
These functionals give total correlation
energies that are around 17--20\% from reference coupled-cluster CCSD(T) values, just around a factor of two worse than
M{\o}ller-Plesset MP2 calculations\cite{usJCP}. Furthermore, atomization energies are basically of the same quality as DFT using
the B3LYP functional\cite{usJCP}.
Other properties of molecular systems like ionization potentials,~\cite{pernalip,leiva,piris_theo,piris_os} 
the chemical hardness,~\cite{our_gap,piris_os} 
dipole moments and static polarizabilities,~\cite{pernal_barends,piris,piris_os} and
vibrational frequencies.~\cite{leiva} 
were also calculated with these functionals with very promising results.

In this Article, we propose a fully empirical form --- that we will denote by ``ML'' --- for the function $f(n_{j},
n_{k})$.  This approach, of using completely empirical functionals, has an already long history\cite{DFTfunc} in
standard DFT, and yields some of the best functionals to date\cite{DFTfit}.

We write $f(n_{j},n_{k})$ as a general Padé approximant
\begin{equation}
  f^{\rm ML }(n_j,n_k) = x \frac{a_0 + a_1 x + a_2 x^2 + \cdots}{1 + b_1 x + b_2
    x^2+\cdots}
  \,,
\end{equation}
where $x=n_j n_k$. We chose a Padé approximant as
it is simple and one of the best choices for approximating a rational function.
Note that we multiplied the Padé by $x$ to ensure that the contribution of
completely empty states ($n_j = 0$) to the exchange-correlation energy is
zero. Furthermore, to recover the Hartree-Fock limit in the case of an
idempotent density matrix, we force $f^{\rm ML}(1) = 1$ by setting
\begin{equation}
  \label{eq:a0}
  a_0 = 1 + \sum_{i=1}^{m} (b_i - a_i)
  \,.
\end{equation}
We also investigated the effects of removing the SI terms from the
functionals (as suggested, e.g., in Ref.\onlinecite{GU}), by trying
the self-interaction corrected (SIC) functional
\begin{multline}
  f^{\rm ML-SIC}(n_j,n_k) = f^{\rm ML}(n_j
  n_k) \\
  - \delta_{jk} \left[f^{\rm ML}(n_j^2) - n_j^2 \right]
\end{multline}
No other constraints are imposed on the functional. This functional form was
then optimized by a downhill simplex method, using as objective function
\begin{equation}
  \label{eq:bardelta}
  \bar \delta = 100 \sqrt{\frac{1}{N} \sum_{i=1}^{N} 
    \left(\frac{E_\text{c}^\text{ML} -
        E_\text{c}^\text{ref}}{E_c^\text{ref}}\right)^2 
  }
  \,,
\end{equation}
i.e., the average percentile error of the correlation energy for a test set of
$N$ molecules. For the reference correlation energy, $E_c^\text{ref}$ we used
values computed with accurate coupled cluster theory [CCSD(T)] with the same
basis set.

The optimization of the Padé was performed in a very pragmatic way. First of
all, we used a small number of parameters in the Padé. This reduces the problem
of over-fitting of the functional, and simplifies the minimization procedure by
keeping the dimensionality of the search space small. Furthermore, we used the
Cartesian 6-31G$^*$ Gaussian basis-set. This basis set is relatively small,
allowing us to quickly perform the many calculations needed to optimize the
functional. The optimization was performed in two steps: i)~The functional was minimized for
a test set consisting on the eight of the smallest molecules of the G2 set\cite{g2set},
namely H$_2$, hydrogen fluoride, lithium hydride, water, ammonia, hydrogen
chloride, hydrogen sulfide, and methane. ii)~We then refined the Padé in the
test set consisting of all closed-shell molecules in the G2-1 set (plus H$_2$).
This set includes 35 molecules, with correlation energies ranging from around
-0.02\,Hartree (LiH) to -0.5\,Hartree (CO$_2$).

\begin{table}[t!]
  \setlength{\tabcolsep}{0.2truecm}
  \caption{Parameters used to fit the functionals. Two versions are given, one
    where the self-interaction terms are explicitly excluded from the
    functionals, and another where they are not. Note that there are only two
    fitted parameters ($a_1$ and $b_1$), as $a_0$ is obtained through
    Eq.~\eqref{eq:a0}.}
  \label{table:fitparams}
  \begin{center}
    \begin{tabular}{lccc}
      & $a_0$ & $a_1$ & $b1$ \\
      \hline \\[-2mm]
      ML     & 126.3101 & 2213.33 & 2338.64 \\
      ML-SIC & 1298.780 & 35114.4 & 36412.2
    \end{tabular}
  \end{center}
\end{table}

Our best functionals are summarized in Table~\ref{table:fitparams}. Note that
these functionals only depend on two parameters! In fact, we did not succeed in
ameliorating the functional significantly by increasing their number. All the
improvements were marginal (of less than 0.5\%), and very often not consistent
(improving the correlation energy of some subset of molecules but worsening
others). We believe, in fact, that depending on just two parameters is a
strength of our functionals, that in this way ally precision with simplicity.

\begin{table}[t]
  \setlength{\tabcolsep}{0.1truecm}
  \caption{Error in the correlation energies for the closed-shell molecules of
    the G2 set calculated using the 6-31G$^*$ basis set. The values in the second
    and third columns are in atomic units. See text for an explanation of the
    meaning of the columns.
  }
  \label{table:res:631G}
  \begin{center}
    \begin{tabular}{cllrl} 
      Method
      & \multicolumn{1}{c}{$\bar \Delta$} 
      & \multicolumn{1}{c}{$\Delta_{\rm max}$}
      & \multicolumn{1}{c}{$\bar \delta$} 
      & \multicolumn{1}{c}{$\delta_{\rm max}$}
      \\
      \hline \\[-2mm]
      M\"uller     & 0.58  & 1.23 (C$_2$Cl$_4$)  & 128.6\% & 438.3\% (Na$_2$) \\
      GU           & 0.28  & 0.79 (C$_2$Cl$_4$)  & 52.14\% & 102.8\% (Na$_2$) \\
      CGA          & 0.23  & 0.55 (C$_2$Cl$_4$)  & 63.92\% & 332.0\% (Na$_2$) \\
      BBC1         & 0.31  & 0.75 (C$_2$Cl$_4$)  & 65.27\% & 159.1\% (Na$_2$) \\
      BBC2         & 0.20  & 0.50 (C$_2$Cl$_4$)  & 44.25\% & 125.0\% (Na$_2$) \\
      BBC3         & 0.074 & 0.27 (C$_2$Cl$_4$)  & 17.84\% & $\phantom{1}$49.0\% (SiH$_2$)\\
      PNOF0        & 0.078 & 0.32 (SiCl$_4$)     & 16.97\% & $\phantom{1}$46.0\% (Cl$_2$) \\
      PNOF         & 0.12  & 0.42 (SiCl$_4$)     & 22.38\% & $\phantom{1}$77.3\% (F$_2$) \\
      {\bf ML}     & 0.059 & 0.18 (pyridine)     & 11.02\% & $\phantom{1}$35.7\% (Na$_2$)\\
      {\bf ML-SIC} & 0.062 & 0.21 (pyridine)     & 10.69\% & $\phantom{1}$42.9\% (Na$_2$) \\
      \hline \\[-2mm]
      MP2         &  0.042 & 0.074 (C$_2$Cl$_4$) & 10.97\% & $\phantom{1}$35.7\% (Li$_2$) \\      
    \end{tabular}
  \end{center}
\end{table}

\begin{table}[t]
  \setlength{\tabcolsep}{0.1truecm}
  \caption{Error in the correlation energies for the closed-shell molecules of
    the G2-1 set calculated using the cc-pVDZ basis set. The values in the 
    second and third columns are in atomic units. See text for an explanation of the
    meaning of the columns.
  }
  \label{table:res:pVDZ}
  \begin{center}
    \begin{tabular}{cllrl} 
      Method
      & \multicolumn{1}{c}{$\bar \Delta$} 
      & \multicolumn{1}{c}{$\Delta_{\rm max}$}
      & \multicolumn{1}{c}{$\bar \delta$} 
      & \multicolumn{1}{c}{$\delta_{\rm max}$}
      \\
      \hline \\[-2mm]
      M\"uller     & 0.35  & 0.57 (Cl$_2$) &  155\% & 439\% (Na$_2$) \\
      GU           & 0.13  & 0.28 (Cl$_2$) & 46.4\% & 120\% (Na$_2$) \\
      CGA          & 0.16  & 0.32 (S$_2$)  & 89.9\% & 331\% (Na$_2$) \\
      BBC1         & 0.19  & 0.35 (Cl$_2$) & 72.8\% & 181\% (Na$_2$) \\
      BBC2         & 0.12  & 0.24 (Cl$_2$) & 51.2\% & 145\% (Na$_2$) \\
      BBC3         & 0.048 & 0.14 (Cl$_2$) & 21.4\% & 68.9\% (Na$_2$) \\
      PNOF0        & 0.045 & 0.14 (Cl$_2$) & 18.0\% & 44.6\% (Na$_2$) \\
      PNOF         & 0.052 & 0.16 (Cl$_2$) & 19.7\% & 49.5\% (Cl$_2$) \\
      {\bf ML}     & 0.026 & 0.064 (N$_2$) & 11.1\% & 45.6\% (Na$_2$) \\
      {\bf ML-SIC} & 0.023 & 0.056 (N$_2$) & 11.5\% & 45.5\% (Na$_2$) \\
      \hline \\[-2mm]
      MP2         &  0.042 & 0.074 (C$_2$Cl$_4$) & 10.97\% & $\phantom{1}$35.7\% (Li$_2$) \\      
    \end{tabular}
  \end{center}
\end{table}

We then tested these functionals for all closed-shell molecules in the whole G2 set\cite{g2set} (119 molecules) using
the same 6-31G$^*$ basis-set, and in the G2-1 basis set (35 molecules) using the better cc-pVDZ correlation consistent
Dunning basis set. Results are summarized in Tables~\ref{table:res:631G} and~\ref{table:res:pVDZ}, where we also
included results obtained with M{\o}ller-Plesser MP2 theory, and other of the most known RDMFT functionals, namely:
M\"uller\cite{mueller}; Goedecker and Umrigar (GU)\cite{GU}; Cs\'anyi, Goedecker, and Arias (CGA)\cite{csgoe};
Gritsenko, Pernal, and Baerends (BBC1, BBC2, and BBC3)~\cite{gritsenko}; and the Piris functionals, both with the
correction to avoid pinned states at one (PNOF) and without (PNOF0)\cite{piris}.

The meaning of the columns is the following
\begin{subequations}
\begin{align}
  \bar \Delta & = \sqrt{\frac{1}{N}\sum_{i=1}^N \left(E_{\rm c} - E_{\rm c}^{\rm
      ref}\right)^2}
  \\
  \Delta_{\rm max} & =\max\left|E_{\rm c} - E_{\rm c}^{\rm (ref)}\right|
  \\
  \delta_{\rm
    max} & =100\times\max\left|\frac{E_{\rm c} - E_{\rm c}^{\rm ref}}{E_{ \rm c}^{\rm
      ref}}\right|
  \,.
\end{align}
\end{subequations}
[The quantity $\bar\delta$ is defined as in Eq.~\eqref{eq:bardelta}.]

The improvement of the functionals over the past years is quite remarkable. The
best functionals that we have available at the moment are literally an order of
magnitude better than the seminal M\"uller functional, regardless of the
criterion used to judge them. We can also notice that our ML functionals are clearly the
best RDMFT functionals. We especially remark the average percentile error
$\bar\delta$ that shows an improvement of more than 60\% with respect to the
BBC3 result, being better than even MP2 theory. The maximum errors are,
unfortunately, still of a somewhat lower quality than MP2 theory. Note also that similar
final results are obtained including or excluding the self-interaction terms in the
functional.

\begin{figure}[t]
  \centerline{\includegraphics[width=0.9\columnwidth]{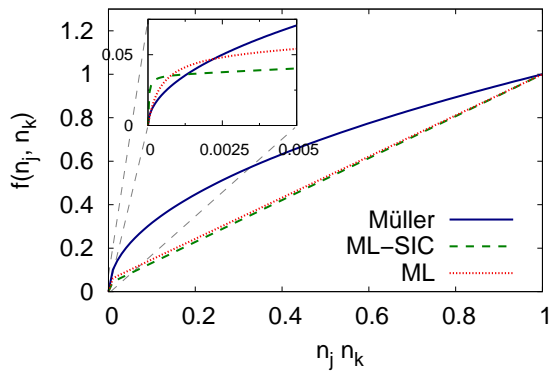}}
  \caption{
    \label{fig:funcs}
    (Color online) Plot of the function $f(n_j,n_k)$ for the ML, ML-SIC and the Müller
    functional. The inset zooms on the particularly interesting region close to
    zero.  }
\end{figure}

In Fig.~\ref{fig:funcs} we plot the ML functionals compared to the simple Müller form. It is clear that, in order to
correct for the consistent over-correlation of the Müller functional, our functionals are almost always smaller than the
simple square-root (with exception of a small region close to zero). We also note that the functionals never become
negative (this was true for all forms that we tried, including those with more parameters). However, the most striking
fact is that the ML functionals are essentially linear except very close to zero, where their value drops quickly to
zero (see inset of Fig.~\ref{fig:funcs}). It is also important to note that all functionals of reasonable quality that
we found exhibited this kind of kink.

\begin{figure}[t]
  \centerline{\includegraphics[width=0.9\columnwidth]{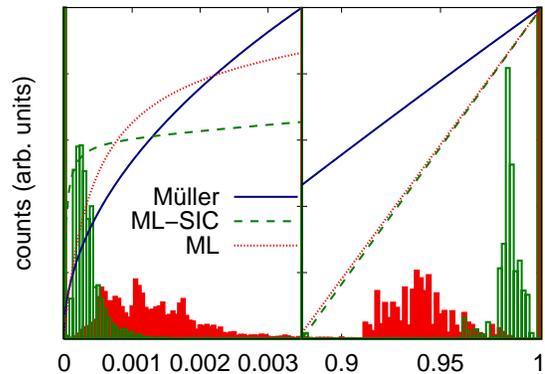}}
  \caption{
    \label{fig:occ}
    (Color online) Histogram with the values of $n_jn_k$ for all G2 closed-shell molecules.  The values of $n_jn_k$ are obtained with
    the both with the ML (dashed red bars) and ML-SIC (empty green bars) functionals.  Note that the left and right
    panels do not have the same horizontal or vertical scale.}
\end{figure}

To better understand our numerical results, we show, in Fig.~\ref{fig:occ}, a histogram of the product $n_jn_k$ for all
G2 closed-shell molecules, calculated both with the ML and ML-SIC functional. We can divide the product of the states in
4 different sets (we will use the nomenclature of Ref.~\onlinecite{gritsenko}):
\begin{enumerate}
  \item Products of weekly occupied orbitals (the large bar at zero). Note that in all cases, these
    weekly occupied orbitals are not pinned at zero, but instead have a very small but finite 
    occupation.
  \item Products of a weekly occupied orbital with a strongly occupied orbital. These are the
    products whose values lie mainly between zero and $\sim 3\times 10^{-3}$, but with a long
    tail that extends to $n_jn_k=0.25$ (ML) and $n_jn_k=0.15$ (ML-SIC).
  \item Products of two strongly occupied orbitals. These have a tail that extends from 
    $n_jn_k=0.83$ (ML) and $n_jn_k=0.95$ (ML-SIC) to one.
  \item Products of two pinned states at one. For curiosity, we refer that both the ML and ML-SIC
    functionals yield exactly the same number of pinned orbitals for the molecules considered.
\end{enumerate}
The ML functional lead to occupation numbers (and corresponding products $n_jn_k$) with a much
broader distribution than ML-SIC. This is not surprising, as the width of the distribution can
be seen as a measure of ``correlation''. In the ML-SIC functional, a large contribution of the
exchange-correlation energy (the self-interaction) has been explicitly subtracted, so the 
resulting occupation numbers will seem less correlated. We stress, however, that from the point
of view of the total correlation energy both ML and ML-SIC fare equally well, even if it 
apparently easier to construct SIC functionals.

The analysis of the picture also makes clear a path to improve RDMFT functionals: to use 
different functional forms to approximate the exchange-correlation in each of the well-separated ranges 
corresponding to weakly-weakly, weakly-strongly, and strongly-strongly occupied states.
The price to pay is a clear increase of the complexity of the functional. This is, in fact,
the path already used by, for example, the BBC corrections\cite{gritsenko}. 

In conclusion, in this Article we proposed a very simple empirical functional to be used within RDMFT.  This functional
is very precise, reaching in some respects the accuracy of more traditional quantum chemical approaches, like MP2
theory.  Perhaps surprisingly, we can reach the same level of average accuracy using functionals that are explicitly
self-interacted corrected or not. This fact is further analyzed by looking at the basic ingredient of the functionals:
the product $n_jn_k$ between the occupation numbers of two states.

\begin{acknowledgments}
  This work was supported in part by the Deutsche Forschungsgemeischaft within the program SPP 1145, by the Portuguese
  FCT through the project PTDC/FIS/73578/2006, and by the EC Network of Excellence NANOQUANTA (NMP4-CT-2004-500198).
  Calculations were performed at the Laborat\'orio de Computa\c{c}\~ao Avan\c{c}ada of the University of Coimbra.
\end{acknowledgments}


%

\end{document}